
\magnification=1200
\baselineskip=12pt
\overfullrule=0pt
\tolerance=100000
\font\title=cmr8
\hsize = 5.8 truein
\vsize = 8.5 truein
\def\lsim{<\kern-2.5ex\lower0.85ex\hbox{$\sim$}\ }
\def\rsim{>\kern-2.5ex\lower0.85ex\hbox{$\sim$}\ }
\ \ \
\bigskip
\centerline{\bf SUPER TRIPLE SYSTEMS AND APPLICATIONS TO}
\smallskip
\centerline{\bf PARA-STATISTICS AND YANG-BAXTER EQUATION}
\vskip 1 cm
\centerline{S. OKUBO}
\centerline{\it Department of Physics and Astronomy,
 University of Rochester}
\centerline{\it Rochester, NY 14627, USA}
\vskip 1 cm
\centerline{ABSTRACT}
\smallskip
 {\narrower\smallskip\noindent
We introduce the notion of
ortho-symplectic super triple system, and apply
it to find solutions of super Yang-Baxter equation.  Also,
 the para-statistics are formulated as a
Lie-super triple system. \smallskip}
\vskip .5 cm
\baselineskip=14pt
\noindent {\bf 1. Triple Product}
\medskip
\noindent {\it 1.1 Quaternionic and Octonionic Triple Systems}
\smallskip
Let $V$ be a $N$-dimensional vector space, which we write hereafter as
$$N=\  {\rm Dim}\ V \quad . \eqno(1.1)$$
A triple product $w = [x,y,z]$ in $V$ is an assignment of $w\ \epsilon\ V$
for any three elements $x,\ y,\ z\ \epsilon\ V$, which is linear in each
variables $x,\ y,\ {\rm and}\ z$.  In other words, we may write
$$[x,y,z]\ ;\ V \otimes V \otimes V \rightarrow V \quad . \eqno(1.2)$$

Perhaps the simplest non-trivial example is the case of $N=4$, which can be
constructed below.  Let $e_1,\ e_2,\ e_3,\ {\rm and}\ e_4$ be a basis of
$V$ and assign $[e_j, e_k,e_\ell]\ \epsilon\ V$ by
$$\left [ e_j, e_k , e_\ell \right] = \sum^4_{m=1} \epsilon_{jk \ell m} e_m
\eqno(1.3)$$
for any $j,k,\ell = 1,2,3,4$, where $\epsilon_{jk \ell m}$ is the totally
antisymmetric Levi-Civita symbol with $\epsilon_{1234} = 1$ in
4-dimensional space.  We then extend the definition of $[x,y,z]$ for any
$x,\ y,\ z\ \epsilon \ V$ from Eq. (1.3) by the linearity.  Further, we can
introduce a bi-linear form $<x |y>$ in $V$ by
$$<e_j |e_k>\ = \delta_{jk} \quad (j,k = 1,2,3,4) \quad .\eqno(1.4)$$
Then, $<x|y>$ defines clearly a symmetric bi-linear non-degenerate form in
 $V$.  The triple product $[x,y,z]$ with the bi-linear form $<x |y>$ can be
easily verified to satisfy the following properties:

\itemitem{(i)} $\quad <x|y>\ =\ <y|x>$ is non-degenerate, \hfill (1.5a)
\smallskip
\itemitem{(ii)} $\quad [x,y,z]$ is totally antisymmetric in
$x,\ y,\ z$ , \hfill (1.5b)
\smallskip
\itemitem{(iii)} $\quad <w|[x,y,z]>$ is totally antisymmetric in
$w,\ x,\ y,\ z$ , \hfill (1.5c)
\smallskip
\itemitem{(iv)} $\quad <[x,y,z]|[u,v,w]>$
\smallskip

\itemitem{    } $\qquad = \ <x|u><y|v><z|w>\ + \ <x|v><y|w><z|u>$
\smallskip
\itemitem{    } $\ \ \qquad + \ <x|w><y|u><z|v>\  - \ <x|w><y|v><z|u>$
\smallskip
\itemitem{    } $\ \ \qquad -\ <x|v><y|u><z|w>\ -\ <x|u><y|w><z|v>$ ,
\hfill (1.5d)
\smallskip

\itemitem{(v)} $\quad [u,v,[x,y,z]]$
\smallskip

\itemitem{    } $\qquad = (<y|v><z|u>\ -  \ <y|u><z|v>)x$
\smallskip
\itemitem{    } $\ \ \qquad +\  (<z|v><x|u>\  - \ <z|u><x|v>)y$
\smallskip
\itemitem{    } $\ \ \qquad +\  (<x|v><y|u>\ -\ <x|u><y|v>)z$ .
\hfill (1.5e)
\smallskip
\noindent For examples, Eq. (1.5c) is an immediate consequence of
$$<e_m |[e_j, e_k, e_\ell]>\ = \epsilon_{jk \ell m}$$
being totally antisymmetric in $j,\ k,\ \ell, \ {\rm and}\ m$, while
Eqs. (1.5d) and (1.5e) is equivalent to the validity of the identity,
$$\eqalign{\sum^4_{m=1} &\epsilon_{jk \ell m} \epsilon_{abcm} =
\delta_{ja} \big( \delta_{kb} \delta_{\ell c} -
\delta_{kc} \delta_{\ell b} \big)\cr
&+ \delta_{jb} \big( \delta_{kc} \delta_{\ell a} - \delta_{ka}
\delta_{\ell c}\big)  + \delta_{jc} \big( \delta_{ka} \delta_{\ell b}
- \delta_{k b} \delta_{\ell a} \big) \quad .\cr}$$
Conversely, we can prove that any system satisfying Eqs. (1.5) is possible
only for Dim $V=4$, and moreover that we can find a basis $e_1,\
e_2,\ e_3\ {\rm and}\ e_4$ which satisfy Eqs. (1.3) and (1.4).  We call the
system to be a quaternionic triple (or ternary) system$^1$.

We can generalize Eqs. (1.5) by adding extra terms
$$\eqalign{\beta \{ &<x|u><y|[z,v,w]>\cr
&+ \ <y|u><z|[x,v,w]>
+
\ <z|u><x|[y,v,w]>\cr
&+\ <x|v><y|[z,w,u]>
+ \ <y|v><z|[x,w,u]>\cr
& +\  <z|v><x|[y,w,u]> \ +\ <x|w><y|[z,u,v]>\cr
& + \ <y|w><z|[x,u,v]>\ +
\ <z|w><x|[y,u,v]>\} \quad , \cr}\eqno(1.5d)^\prime$$
to the right side of Eq. (1.5d) and
$$\eqalign{-\beta \{ &<u|[v,y,z]>x\  + \ <u|[v,z,x]>y\  +
\ <u|[v,x,y]>z\} \cr
-\beta \{&<x|v>[u,y,z]\  + \ <y|v>[u,z,x]\  +\ <z|v>[u,x,y]\cr
&+\ <x|u>[v,z,y]\  + \ <y|u>[v,x,z]\  +
\ <z|u>[v,y,x]\}\quad ,\cr}\eqno(1.5e)^\prime$$
to the right side of Eq. (1.5e) for a constant $\beta$, while other
relations Eqs. (1.5a), (1.5b), and (1.5c) remain unchanged.  We have shown
elsewhere$^1$ that the new system is possible only for two cases of
\smallskip
\itemitem{(a)} $\quad N=4\quad, \quad \beta =0$\hfill (1.6a)
\smallskip
\itemitem{(b)} $\quad N=8 \quad , \quad
\beta = \pm 1 \quad .$\hfill (1.6b)
\smallskip
\noindent We call the new case of $N =$ Dim $V = 8$ to be an octonionic
triple system.  Let us normalize
 $\beta$ to be $\beta = -1$ for the octonionic triple
system by changing the sign of $[x,y,z]$ if necessary.  Let
$e\ \epsilon\ V$ be any fixed element satisfying
$$<e|e>\ = 1 \quad , \eqno(1.7a)$$
and introduce a bi-linear product $xy$ in $V$ by
$$xy = [x,y,e]\  + \ <x|e>y\  + \ <y|e>x\  - \ <x|y>e \quad . \eqno(1.7b)$$
Then, it is easy to show the validity of
$$\eqalignno{&xe = ex = x \quad , &(1.8a)\cr
&<xy|xy>\ = \ <x|x><y|y> &(1.8b)\cr}$$
so that the bi-linear product $xy$ defines the 8-dimensional octonion
algebra$^2$.  Conversely, we can determine the original triple product
$[x,y,z]$ in terms of the bi-linear octonionic product, although we will
not go into detail.

\medskip

\noindent {\it 1.2 Orthogonal Triple System}

\smallskip

Hereafter in this note, we assume that $<x|y>$ is a bi-linear
non-degenerate form in $V$ which is, however, not necessarily symmetric.
Also, let $e_1,\ e_2,\ \dots,\ e_N$ be a basis of $V$ and set
$$g_{jk} = \ <e_j | e_k> \quad . \eqno(1.9)$$
Because of the non-degeneracy of $<x|y>$, $g_{jk}$ possesses its inverse
$g^{jk}$ satisfying
$$\sum^N_{\ell = 1} g_{j \ell} g^{\ell k} =
\sum^N_{\ell = 1} g^{k \ell} g_{\ell j} = \delta^k_j \quad . \eqno(1.10)$$
We set
$$e^j = \sum^N_{k=1} g^{jk} e_k \quad , \quad e_j = \sum^N_{k=1} g_{jk}
e^k \quad . \eqno(1.11)$$
Then, we can expand any $x\ \epsilon\ V$ by
$$x = \sum^N_{j=1} e_j <e^j |x>\ = \sum^N_{j=1} <x |e_j>e^j
\quad .\eqno(1.12)$$

We can now introduce the notion of orthogonal triple system as follows.
Consider a vector space $V$ with a triple linear product $x\ y\ z$ and with
a bi-linear non-degenerate form $<x|y>$, satisfying the axioms
\smallskip
\itemitem{(i)} $\quad <y|x>\ = \ <x|y>$\hfill (1.13a)
\smallskip
\itemitem{(ii)} $\quad y\ x\ z + x\ y\ z = 0$ \hfill (1.13b)
\smallskip
\itemitem{(iii)} $\quad x\ z\ y + x\ y\ z =
2 \lambda <y |z>x - \lambda <x|y>z
 - \lambda <z|x>y \quad ,$ \hfill (1.13c)
\smallskip
\itemitem{(iv)} $\quad <uvx|y>\ = -<x|uvy> \quad ,$ \hfill (1.13d)
\smallskip
\itemitem{(v)} $\quad uv(xyz) = (uvx)yz + x(uvy) z + xy
(uvz) \quad ,$ \hfill
(1.13e)
\smallskip
\noindent where $\lambda$ is a constant.
  We call any such a $V$ to be an orthogonal
triple system$^1$.

For any orthogonal triple system, we can introduce the second triple
product $[x,y,z]$ by
$$x\ y\ z = [x,y,z] + \lambda <y|z>x - \lambda <z|x>y \eqno(1.14)$$
Then, Eqs. (1.13a)--(1.13d) imply that both $[x,y,z]$ and
$<w|[x,y,z]>$ are totally antisymmetric in $x,\ y,\ {\rm and}\ z$, as well as
in $w,\ x,\ y,\ {\rm and}\ z$, respectively.  However, the last relation
Eq. (1.13e) will become a rather complicated equation in terms of
$[x,y,z]$'s.  The notion of the orthogonal triple system is a
generalization of both quaternionic and octonionic triple systems.  Indeed
for the totally antisymmetric product $[x,y,z]$ of both quaternionic and
octonionic
triple systems, we conversely introduce the triple product $x\ y\ z$ by
Eq. (1.14) where the constant $\lambda$ is assumed to be
\smallskip
\itemitem{(a)} $\quad \lambda =$ arbitrary for $N=4$ \hfill (1.15a)
\smallskip
\itemitem{(b)} $\quad \lambda = - 3\  \beta$ for $N=8\quad .$
\hfill (1.15b)
\smallskip
\noindent We can then verify the validity of Eqs. (1.13), so that they
become orthogonal triple systems.  The reason why we can choose
$\lambda$ to be arbitrary for the case of $N=4$ is validities of special
identities, as has been explained elsewhere$^1$.
\bigskip
\vfill\eject
\noindent {\bf 2. Lie-Super Triple System}
\medskip
\noindent {\it 2.1 Super Space}
\smallskip

Let the underlying vector space $V$ admit a $Z_2$-grading, so that
$$V = V_B \oplus V_F \eqno(2.1)$$
is a direct sum of the bosonic vector space $V_B$ and of the fermionic
vector space $V_F$.  We set their dimensions as
$${\rm Dim}\ V_B = N_B \quad , \quad {\rm Dim}\ V_F = N_F \eqno(2.2)$$
so that
$$N = N_B + N_F \quad . \eqno(2.3)$$
We define the signature function $\sigma (x)$ in $V$ by
$$\qquad \qquad \qquad \qquad\quad \sigma (x) =
 \cases{0\ , &if $x\ \epsilon\ V_B$ \hbox to 2.1in{\hfil(2.4a)}\cr
\noalign{\vskip 7pt}%
1\ , &if $x\ \epsilon\ V_F \quad . \hbox to 2in{\hfil (2.4b)}$\cr}$$

\noindent We consider only homogeneous elements $x$ of $V$, i.e. either
$x\ \epsilon\ V_B$ or $x\ \epsilon \ V_F$.  Any triple product
$x\ y\ z\ ({\rm or}\ [x,y,z])$
 defined in $V$ is hereafter assumed to satisfy
$$\sigma (xyz) = \{\sigma (x) + \sigma (y) + \sigma (z) \}\quad
({\rm mod}\ 2) \quad . \eqno(2.5)$$
Also, any bi-linear non-degenerate form $<x|y>$ in $V$ is called
 super-symmetric, if we have
\smallskip
\itemitem{(a)} $\quad <x|y>\ = 0$, unless $\sigma(x) = \sigma (y)$ (mod 2) ,
\hfill (2.6b)
 \smallskip
\itemitem{(b)} $\quad <y|x>\ = (-1)^{\sigma (x) \sigma (y)} <x|y>$,
\hfill (2.6c)
\smallskip

\noindent which
 we assume throughout in this note.  In what follows, we often write
$$(-1)^{xy} = (-1)^{\sigma (x) \sigma (y)} \eqno(2.7)$$
whenever there is no confusion.
\medskip

\noindent {\it 2.2 Lie-Super Triple System}

\smallskip

Suppose that a triple product $x\ y\ z$ satisfies the following axioms.
\smallskip
\itemitem{(i)} $\quad y\ x\ z = -(-1)^{xy} x\ y\ z$,\hfill (2.8a)
\smallskip
\itemitem{(ii)} $\quad (-1)^{xz} x\ y\ z + (-1)^{yx} y\ z\ x +
(-1)^{zy} z\ x\ y = 0$, \hfill (2.8b)
\smallskip
\itemitem{(iii)} $\quad uv (xyz) = (uvx)yz + (-1)^{(u+v)x} x(uvy) z$
\smallskip
\itemitem{    }  $\qquad \quad + (-1)^{(u+v)(x+y)} xy (uvz)$ \hfill (2.8c)
\smallskip
\noindent where we have set for simplicity
$$(-1)^{(u+v)x} = (-1)^{ux +vx} =
(-1)^{[\sigma (u) +
\sigma (v)] \sigma (x)}$$
etc. in accordance with Eq. (2.7).  If $V = V_B$ with $V_F = 0$, then this
 reduces to the well-known Lie triple system$^{3,4}$, so that we call the
system satisfying Eqs. (2.8) to be a Lie-super triple system.

The Lie-super triple system is intimately related to Lie-super algebra$^5$.
 Let $L$ be a Lie-super algebra with the Lie-product $[x,y]$, so that we
have
\smallskip
\itemitem{(i)} $\quad \sigma ([x,y]) = \{ \sigma (x)
+ \sigma (y)\}$ mod 2 \hfill (2.9a)
\smallskip
\itemitem{(ii)} $\quad [y,x] = -(-1)^{xy} [x,y]$, \hfill (2.9b)
\smallskip
\itemitem{(iii)} $\quad (-1)^{xz} [[x,y],z] +
(-1)^{yx} [[y,z],x] + (-1)^{zy}
[[z,x],y] = 0$ . \hfill (2.9c)
\smallskip
\noindent If we set
$$x\ y\ z \equiv [[x,y],z] \quad , \eqno(2.10)$$
then it is not difficult to see that the triple product $x\ y\ z$ defined
 by Eq. (2.10) satisfies Eqs. (2.8).  The converse statement is also true
in a sense to be specified below.

For this, we introduce first the linear multiplication operator
$L_{x,y}$ in $V$ by
$$L_{x,y} z = x\ y\ z \quad , \eqno(2.11)$$
and set
$$M =\ {\rm vector\ space\ spanned\ by}\
 L_{x,y} \hbox{'}{\rm s}\ ,\
(x,\ y\  \epsilon\ V) \quad .
\eqno(2.12)$$
Calculating the commutators
$$[L_{x,y},L_{u,v} ] = L_{x,y} L_{u,v} - (-1)^{(x+y)(u+v)}
L_{u,v} L_{x,y}
\eqno(2.13)$$
\noindent from Eq. (2.8c), we find
$$[L_{x,y}, L_{u,v}] = L_{xyu,v} +
(-1)^{(x+y)u} L_{u,xyv} \eqno(2.14)$$
so that this defines a Lie-super algebra with
$$\sigma (L_{x,y}) = [\sigma (x) + \sigma (y) ]\ ({\rm mod}\ 2)
\eqno(2.15)$$
Actually, we can construct a larger Lie-super algebra in a larger space
$$V_0 = V \oplus M \eqno(2.16)$$
as follows.  The commutators $[M,M]$ are still defined by Eq. (2.13), while
we set
$$[L_{x,y},z] = - (-1)^{(x+y)z} [z, L_{x,y}] \equiv x\ y\ z \eqno(2.17)$$
for $[M,V]$ and
$$[x,y] = -(-1)^{yx} [y,x] \equiv L_{x,y} \eqno(2.18)$$
for $[V,V]$.  We note especially
$$[M,M] \subset M \quad , \quad [M,V]\subset V \quad , \quad [V,V]\subset
 M \quad ,
\eqno(2.19)$$
which is familiar in the theory of symmetric homogeneous spaces.  It is not
hard to verify that these define a Lie-super algebra in $V_0$.  If $V = V_B$
with $V_F = 0$, this construction reduces of course to the familiar
canonical construction$^{3,4}$.

Then, we calculate
$$[[x,y],z] = [L_{x,y},z] = x\ y\ z \eqno(2.20)$$
\noindent from Eqs. (2.18) and
 (2.17), so that the relation Eq. (2.10) is formally
valid, although the meaning of $[x,y]$ is different, since it is an element
of $M$, but not of $V$ itself.

\medskip

\noindent {\it 2.3 Para-statistics as a Lie-Super Triple System}

\smallskip

Let $<x_1,x_2 \dots, x_N>$ be a vector space spanned by
$x_1,x_2,\dots,x_N$, and set

$$\eqalignno{V_B &= \ <a_j, a^+_j \ ;\ j = 1,2,\dots, n> &(2.21a)\cr
\noalign{\vskip 4pt}%
V_F &= \ <b_\alpha, b^+_\alpha \ ;\ \alpha = 1,2,\dots, m> \quad .
&(2.21b)\cr}$$
Moreover, we introduce $<x|y>$ in $V = V_B \oplus V_F$ by
$$\eqalignno{<a_j |a^+_k>\ &= \ <a^+_k |a_j>\ = \delta_{jk} \quad ,
\quad (j,k = 1,2,\dots,n) \quad , &(2.22a)\cr
\noalign{\vskip 4pt}%
<b_\alpha |b^+_\beta>\ &=\ -<b^+_\beta |b_\alpha>\ =\ \delta_{\alpha \beta}
\quad , \quad (\alpha , \beta = 1,2,\dots, m)\quad , &(2.22b)\cr}$$
while all other combinations such as $<a_j
|a_k>\ ,\ <a_j|b_\alpha>\ ,\ <a_j|b^+_\alpha>$ etc. are assumed to be
identically zero.  Then, $<x|y>$ defines a bi-linear symmetric
non-degenerate form in the sense of Eqs. (2.6).

For any vector space $V$ with $<x|y>$ satisfying Eq. (2.6), the triple
product defined by
$$x\ y\ z = \lambda <y|z> x -  \lambda (-1)^{yz} <x |z>y \quad , \eqno(2.23)$$
for a non-zero constant $\lambda$ can be readily verified to give a
Lie-super triple system.  Applying this fact to the system $V = V_B
\oplus V_F$ given by Eqs. (2.21) and (2.22), we see that $
V$ is a Lie-super triple system, while $V_B$ itself defines a Lie triple
system.  Especially, for $V = V_B$, we see
$$\eqalignno{[a_j, a^+_k , a_\ell ] &= [[a_j , a^+_k], a_\ell] =
2\ \delta_{k \ell} a_j \quad , &(2.23a)\cr
\noalign{\vskip 4pt}%
[a_j^+, a^+_k , a_\ell ] &= [[a_j^+ , a^+_k], a_\ell] =
2\ \delta_{k \ell} a_j^+ -
2 \ \delta_{j \ell} a^+_k \quad , &(2.23b)\cr}$$
where we have set $\lambda = 2$.  These relations are essentially
equivalent to the para-Fermi statistics$^6$.  Similarly, $V_F$ defines the
para-boson statistics$^6$.  However, for $V = V_B \oplus V_F$, we may have
rather peculiar relations such as
$$[[a_j, b^+_\alpha],a^+_k] = - 2\ \delta_{jk} b^+_\alpha \eqno(2.23c)$$
so that bosonic and fermionic operators are no longer independent of each
other in contrast to the standard theory.  This possibility will be
explored, however, elsewhere.  At any rate, the para-statistic relations
may be rewritten as
$$\eqalignno{&[[x,y],z] = 2<y|z>x - 2\ (-1)^{yz} <x|z>y \quad ,
&(2.24a)\cr
\noalign{\vskip 4pt}%
&[x,y] = xy - (-1)^{xy} yx &(2.24b)\cr}$$
where boson variables and fermion variables are now regarded to possess
opposite signature values as in Eqs. (2.21) and (2.22), i.e.
$$\sigma (x) = \cases{0 &for fermions\cr
1 &for bosons\cr}\eqno(2.25)$$
Also, this realizes a ortho-symplectic Lie-super algebra$^5$.
\bigskip
\noindent {\bf 3. Ortho-Symplectic Super-Triple System}
\medskip
\noindent {\it 3.1 Definition}
\smallskip

Let $<x|y>$ be a bi-linear non-degenerate super-symmetric form in the sense of
Eqs. (2.6).  Suppose now that a
 triple linear product $x\ y\ z$ satisfies the following axioms:

\smallskip

\itemitem{(i)} $\quad <y|x>\ = (-1)^{xy} <x|y>$ , \hfill (3.1a)
\smallskip
\itemitem{(ii)} $\quad x\ y\ z + (-1)^{xy} y\ x\ z = 0$ , \hfill (3.1b)
\smallskip
\itemitem{(iii)} $\quad x\ y\ z + (-1)^{yz} x\ z\ y$
\smallskip
\itemitem{   } $\qquad\qquad = 2\  \lambda <y
|z>x - \lambda <x|y>z - \lambda (-1)^{yz}
<x|z>y$, \hfill (3.1c)
\smallskip
\itemitem{(iv)} $\quad <uvx|y>\ = - (-1)^{(u+v)x} <x|uvy>$ , \hfill (3.1d)
\smallskip
\itemitem{(v)} $\quad uv (xyz) = (uvx)yz + (-1)^{(u+v)x} x(uvy)z$
\smallskip
\itemitem{  } $\qquad \qquad
 +\  (-1)^{(u+v)(x+y)}xy (uvz)$, \hfill (3.1e)

\smallskip

\noindent for a constant $\lambda$.  We call the system to be
ortho-symplectic super-triple system.  If $V = V_B$ with $V_F = 0$,
then it reduces to the orthogonal triple system,
 while the other case of $V = V_F$ with $V_B = 0$ is equivalent to the
symplectic triple system of Yamaguchi and Asano$^7$.  Although the present
super triple system is still a special case of more general super systems
considered by many authors, it is still of some intrinsic interest as we
will see below.

Here, we will construct a class of ortho-symplectic super triple system as
follows.  Let $<x|y>$ be as before, and let
$$J_\alpha \ :\ V \rightarrow V\quad (\alpha = 1,2,\dots,n) \eqno(3.2)$$
be a signature-preserving linear mapping in $V$ so that
$$\sigma (J_\alpha x) = \sigma(x) \quad . \eqno(3.3)$$
Moreover, we assume the validity of
$$<x|J_\alpha y>\ = -<J_\alpha x|y> \eqno(3.4)$$
as well as
$$J_\alpha J_\beta = \lambda \delta_{\alpha \beta} 1 +
\sum^n_{\gamma =1} f_{\alpha \beta \gamma} J_\gamma
\eqno(3.5)$$
for a constant $\lambda$, where 1 stands for the identity linear mapping in
$V$.  We consider two cases of
\smallskip

\itemitem{(i)} $\quad n = 1 \quad , \quad f_{\alpha \beta \gamma} = 0$,
\hfill (3.6a)

\smallskip

\itemitem{(ii)} $\quad n=3 \quad , \quad f_{\alpha \beta \gamma} =
\epsilon_{\alpha \beta \gamma} =$ Levi-Civita symbol  \hfill (3.6b)

\smallskip

\noindent so that the case of $n=3$ corresponds to the quaternion algebra
for $J_\alpha \ (\alpha = 1,2,3)$.  Now, we introduce a triple product by
$$\eqalign{x\ y\ z = \sum^n_{\alpha =1} \big\{ &J_\alpha x
<y|J_\alpha z>\cr
&+\  (-1)^{z(y+x)} J_\alpha y
<z|J_\alpha x>
+ J_\alpha z<x|J_\alpha y>\big\}\cr
& + \lambda \{ <y|z>x - (-1)^{z(x+y)} <z|x>y\} \quad .\cr}\eqno(3.7)$$
We can verify then that it satisfies Eqs. (3.1).

\medskip
\noindent {\it 3.2 Its Relationship to Lie-super Triple System}
\smallskip

Then, from a given ortho-symplectic super-triple system $V$, we can construct
Lie-super triple systems in two different ways.  The first one is
essentially a straight-forward generalization of the method$^7$ by
Freudenthal-Yamaguchi-Asano in the space $V \oplus V$, which will be
reported elsewhere.  Here, we
 will consider a generalization of the earlier
result given in ref. 8 which will be referred hereafter to as II.  Let
$e_1,e_2,\dots, e_N$ be a basis of $V$ and define its dual $e^j$ by
Eq. (1.11).  We introduce a new triple product in the same vector space by
$$x \cdot y \cdot z = - \sum^N_{j=1} (xy e_j) e^j z - {1 \over 3}\
\lambda \left( N_0 - 16 \right) x\ y\ z \quad , \eqno(3.8)$$
where we have set
$$N_0 = N_B - N_F \quad . \eqno(3.9)$$
Then, we can show as in II that $x \cdot y \cdot z$ defines a Lie-super
triple system.  However, it could happen that we have $x \cdot
 y \cdot z = 0$
 identically as in octonionic triple system as well as in other cases
studied in II.  As a matter of fact, the condition
$x \cdot y\cdot z = 0$
 is crucial to obtain some class of solutions  of
 Yang-Baxter equation as we already noted in II.

We now claculate $x \cdot y \cdot z$ for the ortho-symplectic
super-triple system given by Eq. (3.7) to find
$$x \cdot y \cdot z = c\ x*y*z \quad , \eqno(3.10a)$$
$$\qquad\qquad\qquad\qquad\qquad c =
 \cases{-{1 \over 3}\ \lambda (N_0 - 4)\ , &for $n=1$
 \hbox to 1.41in {\hfil (3.10b)}\cr
\noalign{\vskip 7pt}%
-{1 \over 3}\ \lambda (N_0 + 8)\ , &for $n=3$
 \hbox to 1.41in {\hfil (3.10c)}\cr}$$
where $x*y*z$ is a Lie-super triple product defined by
$$\eqalign{x*y*z = \sum^n_{\alpha =1} \big\{ &J_\alpha x
<y|J_\alpha z>\cr
& + (-1)^{z(x+y)} J_\alpha y
<z|J_\alpha x>
-\  2\  J_\alpha z<x|J_\alpha y>\big\}\cr
\noalign{\vskip 4pt}%
& + \lambda \big\{
<y|z>x - (-1)^{z(x+y)} <z|x>y \big\} \quad .\cr}\eqno(3.11)$$
 Especially, we see
$$x \cdot y \cdot z = 0 \eqno(3.12)$$
for two cases of
\smallskip

\itemitem{(i)} $\quad n = 1 \quad , \quad N_0 = 4$ \hfill (3.13a)

\smallskip

\itemitem{(ii)} $\quad n = 3 \quad , \quad N_0 = -8$ . \hfill (3.13b)
\smallskip

We also note that a special case of
$$x\ y\ z = \lambda\  \left\{ <y|z>x - (-1)^{z(x+y)} <z|x>y \right\}
\eqno(3.14a)$$
gives
$$x \cdot y \cdot z = - {1 \over 3}\ \lambda \ \left( N_0 - 10 \right)
 x\ y\ z \quad . \eqno(3.14b)$$
Especially $x\ y\ z$ defined by Eq. (3.14a) describes both ortho-symplectic
 and Lie super triple systems at the same time.

\bigskip
\noindent {\bf 4. Yang-Baxter Equation}
\medskip

Let $R^{cd}_{ab}(\theta)\ (a,b,c,d = 1,2,\dots, N)$ be the scattering
matrix for $a+b \rightarrow c+d$ in one-dimensional line with the rapidity
difference $\theta$.  The (super) Yang-Baxter equation$^9$ is the relation.
$$\eqalign{\sum^N_{a^\prime, b^\prime, c^\prime = 1}
 &(-1)^{a^\prime b^\prime + a_2 c^\prime + b_2 c_2}
R^{b^\prime a^\prime}_{a_1 b_1} (\theta) R^{c^\prime a_2}_{a^\prime c_1}
(\theta^\prime) R^{c_2 b_2}_{b^\prime c^\prime} (\theta^{\prime \prime})\cr
&= \sum^N_{a^\prime , b^\prime , c^\prime = 1} (-1)^{b^\prime c^\prime
+ a^\prime c_2 + a_2 b_2} R^{c^\prime b^\prime}_{b_1 c_1}
(\theta^{\prime \prime}) R^{c_2 a^\prime}_{a_1 c^\prime} (\theta^\prime)
R^{b_2 a_2}_{a^\prime b^\prime} (\theta)\cr}\eqno(4.1)$$
with the energy-momentum conservation law
$$\theta^\prime = \theta + \theta^{\prime \prime} \quad . \eqno(4.2)$$
This relation can be graphically
 depicted as in Fig. 1, below.

\vskip 3in

\centerline{Figure 1}

\medskip

\noindent Here, $(-1)^{ab} = (-1)^{\sigma (a) \sigma (b)}$ is the signature
factor as before.

If we set further
$$S^{cd}_{ab} (\theta) = (-1)^{cd} R^{cd}_{ab} (\theta)
\eqno(4.3)$$
then we can eliminate all sign factors from Eq. (4.1) in terms of
$S^{cd}_{ab} (\theta)$.  However, it will lose then some interesting
features of the super-space.

We assume hereafter that $R^{cd}_{ab} (\theta)$ satisfies a condition
$$R^{cd}_{ab} (\theta) = 0 \quad {\rm if}\quad
 \sigma (c) + \sigma (d) \not= \{
\sigma (a) + \sigma (b) \} \ \ {\rm mod}\ 2 \quad . \eqno(4.4)$$
We now introduce two $\theta$-dependent triple linear products by
$$\eqalignno{\big[ e^b , e_c , e_d \big]_\theta &=
\sum^N_{a=1} e_a R^{ab}_{cd} (\theta) \quad , &(4.5a)\cr
\big[ e^a , e_d , e_c \big]^*_\theta &= (-1)^{ab + cd}
\sum^N_{b=1} e_b R^{ab}_{cd} (\theta) &(4.5b) \cr}$$
or equivalently by
$$\eqalign{R^{ab}_{cd} (\theta) &= \ <e^a | \big[ e^b , e_c , e_d
\big]_\theta>\cr
\noalign{\vskip 4pt}%
&= (-1)^{ab + cd} <e^b | \big[ e^a , e_d , e_c \big]^*_\theta >
\quad . \cr}\eqno(4.6)$$
Note that Eqs. (4.4) and (4.5) are consistent with Eq. (2.5).

Now, the Yang-Baxter equation Eq. (4.1) can be rewritten as a triple
product equation
$$\eqalign{&\sum^N_{j=1} (-1)^{\sigma (e_j) \sigma (z)}
 \big[ v, \big[ u,e_j,z \big]_{\theta^\prime}, \big[ e^j, x, y \big]_\theta
\big]^*_{\theta^{\prime \prime}}\cr
&= (-1)^{\sigma (u) \sigma (v) + \sigma (x) \sigma (z)} \sum^N_{j=1}
(-1)^{\sigma (e_j) \sigma (x)}
\big[ u,\big[ v, e_j, x \big]^*_{\theta^\prime} ,
\big[e^j , z, y \big]^*_{\theta^{\prime \prime}} \big]_\theta \quad .
\cr}\eqno(4.7)$$

Here, we will present a class of solutions of Eq. (4.7).  Let
$x\ y\ z$ be the ortho-symplectic
 super-triple product defined as in section 3.  We seek a solution with
 ansatz of
$$\eqalign{[z,x,y]_\theta = &P(\theta) (-1)^{(x+y)z} x\ y\ z + Q (\theta)
<x|y> z\cr
\noalign{\vskip 4pt}%
&+ R(\theta) <z|x>y + S(\theta) (-1)^{xy} <z|y>x\cr}\eqno(4.8)$$
for some functions $P(\theta), \ Q(\theta),\ R(\theta), \ {\rm and}\
S(\theta)$ of $\theta$ to be determined.  We note that Eq. (4.8) implies
$$[z,x,y]^*_\theta = [z,x,y]_\theta \quad . \eqno(4.9)$$
Moreover, we assume that the triple product $x \cdot y \cdot z$ defined by
Eq. (3.8) vanishes identically, i.e.
$$x \cdot y \cdot z = 0 \quad . \eqno(4.10)$$
Then, the solution for $P(\theta) \not= 0$ can be found to be
$$\eqalignno{R(\theta) / P(\theta) &= a + k \theta \quad , &(4.11a)\cr
\noalign{\vskip 4pt}%
Q (\theta) / P (\theta) &= \lambda - {2 a \lambda \over
 2(a-\lambda ) + k \theta} \quad , &(4.11b)\cr
\noalign{\vskip 4pt}%
S(\theta) / P (\theta ) &= -2\  \lambda - {2 \lambda a \over
k \theta} \quad , &(4.11c)\cr}$$
where we have set
$$a = - {1 \over 6}\ \lambda \ \big( N_0 - 4 \big) = - {1 \over 6} \ \lambda \
\big( N_B - N_F - 4 \big) \quad , \eqno(4.12)$$
and $k$ is an arbitrary constant.
Evidently, the present solution generalizes that given in II.  Especially,
this reproduces the solution of de Vega and Nicolai$^{10}$, which is based
upon the octonionic triple product corresponding to $N_B = 8$ and
$N_F = 0$.  Also, $x\ y\ z$ given by Eq. (3.7) with $n=1$ satisfies
$x \cdot y \cdot z = 0$ for $N_0 =4$ with $a=0$, so that the solution
admits existence of non-zero $P(0),\ Q(0),\ R(0), \ {\rm and}\ S(0)$
 for $\theta = 0$ which may be of some interest to the knot theory.

For other applications of super-triple systems, see also ref. 11.

\bigskip
\noindent {\bf Acknowledgements}
\medskip

This paper is dedicated to the 65th birthday of Professor K. C. Wali of
Syracuse University.  The author wishes a long productive life ahead for
Professor Wali.

This paper is supported in part by the U.S. Department of Energy Grant
No. DE-FG-02-91ER40685.

\bigskip
\noindent {\bf References}
\medskip

\item{1.} S. Okubo, {\it Jour. Math. Phys.} in press.

\item{2.} R. D. Schafer, {\it An Introduction to Non-associative Algebras}
(Academic, New York, 1966).

\item{3.} W. G. Lister, {\it Am. J. Math.} {\bf89} (1952), 787.

\item{4.} K. Yamaguchi, {\it J. Sci. Hiroshima University} {\bf A21},
(1958) 155.

\item{5.} M. Scheunert, {\it The Theory of Lie Superalgebra}
(Springer-Verlag, Berlin-Heidelberg-New York, 1979).

\item{6.} Y. Ohnuki and S. Kamefuchi, {\it Quantum Field Theory
 and Parastatistics} (University of Tokyo Press, Tokyo/Springer-Verlag,
 Berlin, 1982).

\item{7.} K. Yamaguchi and H. Asano, {\it Proc. Jap. Acad.} {\bf 51},
(1972) 247.

\item{8.} S. Okubo, {\it Jour. Math. Phys.} in press.

\item{9.} {\it Yang-Baxter Equation in Integrable Systems}, edited by M.
Jimbo (World Scientific, Singapore, 1989).

\item{10.} H. J. de Vega and H. Nicolai, {\it Phys. Lett.} {\bf B244},
(1990) 295.

\item{11.} M. G\"unaydin, {\it Phys. Lett.} {\bf B255}, (1991) 46.
M. G\"unaydin and S. Hyun, {\it Mod. Phys. Lett.} {\bf 6}, (1991) 1733
and {\it Nucl. Phys.} {\bf B373}, (1992) 688.

\end